\documentclass[12pt]{article} 

\usepackage{hyperref}
\usepackage{url}
\usepackage{setspace}
\usepackage{amsmath,amssymb,amscd,braket}
\usepackage{amsfonts}
\usepackage{color}
\usepackage{graphicx}
\usepackage{cite}

 \def\p{\partial}
 \newcommand{\bea}{\begin{eqnarray}}
\newcommand{\eea}{\end{eqnarray}}
\newcommand{\be}{\begin{equation}}
\newcommand{\ee}{\end{equation}}
\newcommand{\ba}{\begin{align}}
\newcommand{\ea}{\end{align}}

\newcommand\rref[1]{(\ref{#1})}

\newlength{\slength}
\renewcommand{\title}[1]{\vbox{\center\LARGE{#1}}\vspace{5mm}}
\renewcommand{\author}[1]{\vbox{\center#1}\vspace{5mm}}
\newcommand{\address}[1]{\vbox{\center\footnotesize\em#1}}
\newcommand{\email}[1]{\vbox{\center\footnotesize\tt#1}\vspace{5mm}}

\newtheorem{theorem}{Theorem}

\begin{document}

\begin{titlepage}

\begin{center}

\hfill \\
\hfill \\
\begin{flushright}
\hfill{\tt CERN-TH-2020-125}
\end{flushright}
\vskip 1cm

\title{
From sources to initial data and back again: 

on bulk singularities in Euclidean AdS/CFT
}

\author{Alexandre Belin$^{a}$, Benjamin Withers$^{b}$
}

\address{
${}^a$CERN, Theory Division, 1 Esplanade des Particules, Geneva 23, CH-1211, Switzerland
\\
${}^b$Mathematical Sciences and STAG Research Centre, University of Southampton, Highfield, Southampton SO17 1BJ, UK
}

\email{a.belin@cern.ch, b.s.withers@soton.ac.uk}

\end{center}

\abstract{

A common method to prepare states in AdS/CFT is to perform the Euclidean path integral with sources turned on for single-trace operators. These states can be interpreted as coherent states of the bulk quantum theory associated to Lorentzian initial data on a Cauchy slice. In this paper, we discuss the extent to which arbitrary initial data can be obtained in this way. We show that the initial data must be analytic and define the subset of it that can be prepared by imposing bulk regularity. Turning this around, we show that for generic analytic initial data the corresponding Euclidean section contains singularities coming from delta function sources in the bulk. We propose an interpretation of these singularities as non-perturbative objects in the microscopic theory.

}

\vfill

\end{titlepage}

\eject

\tableofcontents

\section{Introduction}

The AdS/CFT correspondence \cite{Maldacena:1997re} relates theories of quantum gravity in AdS$_{d+1}$ to conformal field theories in $d$ spacetime dimensions. A powerful aspect of this duality is that it establishes that the Hilbert spaces of the two theories are the same, namely that the Hilbert space of quantum gravity is the same as that of the dual CFT. The Hilbert space of a conformal field theory placed on a spatial $S^{d-1}$ is completely understood thanks to the state operator correspondence: there is a one-to-one map between energy eigenstates $\ket{E_i}$  and local operators $O_i(x)$ of the CFT. Theories of quantum gravity that are well described by semi-classical general relativity (or supergravity) at low energies are dual to very exotic CFTs, which possess a large number of degrees of freedom while retaining a sparse spectrum and a large gap to higher spin operators \cite{Heemskerk:2009pn,Afkhami-Jeddi:2016ntf,Meltzer:2017rtf,Belin:2019mnx,Kologlu:2019bco,Hartman:2014oaa,Belin:2016yll}.

Within such theories, it is very natural question to ask the following question: which CFT states describe semi-classical geometries? In some cases, the answer is known. For example, the vacuum of the CFT maps to empty AdS, or an eternal black hole maps to the thermofield-double state \cite{Maldacena:2001kr}. While one of the slogans of AdS/CFT has often been to say that geometries are dual to states of the CFT, this statement is slightly imprecise. A full Lorenztian spacetime really describes the time evolution of a state, while a state itself lives at a given moment of time. In other words, given a state $\ket{\psi}$, one can always obtain the state at a different time by time-evolving with the Hamiltonian, namely by applying the operator $e^{-i H t}$ to the state.

\begin{figure}
\begin{center}
\includegraphics[width=0.4\textwidth]{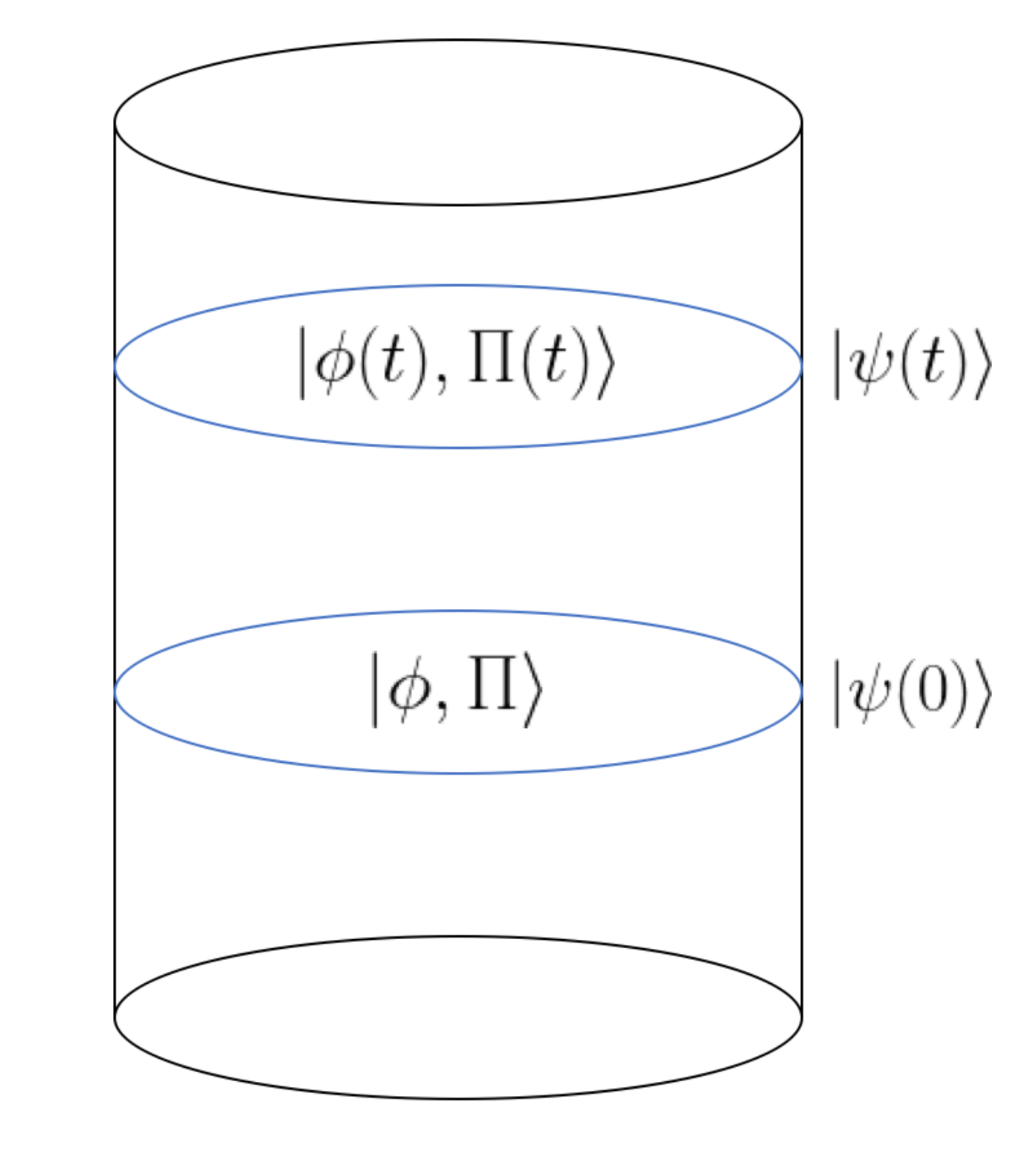}
\end{center}
\caption{A Lorentzian asymptotically AdS geometry. At $t=0$, the CFT state is $\ket{\psi(0)}$ which describes the initial data $\ket{\phi,\Pi}$. The state and initial data can be evolved using the boundary Hamiltonian to some later time $t$. In the CFT, we apply the unitary operator $e^{-iH t}$, and in the bulk, we solve Einstein's equations. At the classical level, this problem is well-posed in the bulk.}
 \label{initialdata}
\end{figure}

On the gravitational side, this statement is equivalent to saying that all that is needed at the semi-classical level to describe a state is Lorentzian initial data, that is data of the gravitational fields on a Cauchy slice $\Sigma$, see Fig. \ref{initialdata}. In a two-derivative theory, initial data is simply given by the value of the field $\phi|_\Sigma$ and its normal derivative $\Pi\sim \p_n \phi|_\Sigma$ on the slice.\footnote{For the gravitational degrees of freedom, this data must satisfy the constraints equations which gives additional restrictions but can be dealt with systematically, see \cite{York:1972sj} or \cite{witten2017,Belin:2018bpg} in the context of holography.} From this data, one can systematically evolve forward in time by solving the equations of motion. In Anti-de Sitter space, boundary conditions at the time-like AdS boundary must also be specified, and the standard Hamiltonian corresponds to turning off all sources at the boundary. At the classical level, it is known in general relativity that this problem is well-posed (see for example \cite{Friedrich:1998xt}).

We will call the states that describe semi-classical geometries $\ket{\phi,\Pi}$ since they are specified by the initial data, and they should be interpreted as coherent states of the full quantum theory. A coherent state behaves as classically as is possible in quantum mechanics and is labelled by a point on phase space, which is precisely what the labels $(\phi,\Pi)$ represent. The question then becomes, which CFT states describe these coherent states
\be
\ket{\psi_{\phi,\Pi}}_{\text{CFT}} \stackrel{?}{=} \ket{\phi,\Pi} \,.
\ee
Naturally, the state operator correspondence guarantees that we can write such a state as
\be \label{energybasis}
\ket{\psi_{\phi,\Pi}}_{\text{CFT}}=\sum_i c_i \ket{E_i} \,.
\ee
The issue with such a description is that one really needs to specify the coefficients $c_i$ for all energy eigenstate of the theory, most of which are complicated. Moreover, the geometry should correspond to some sort of coarse-graining of the exact quantum state and it is thus likely that the precise details of the coefficients $c_i$ are not crucial to the understanding of the state. So while this description is of course correct, it is not very useful in a general context.\footnote{There are some cases where the coefficients and respective states can be identified more clearly, which typically involves supersymmetry. This is the case for the microstate geometries appearing in the Fuzzball program (see for example \cite{Mathur:2005zp,Skenderis:2008qn} reviews) or the LLM geometries \cite{Lin:2004nb}.}

Motivated by holography, a different picture has emerged. Rather than specify the nature of the state itself, it may be convenient to describe instead the way the state is prepared. This has led to the proposal that semi-classical geometries are states prepared by the Euclidean path integral with sources turned on \cite{Skenderis:2008dh,Skenderis:2008dg,Botta-Cantcheff:2015sav,Marolf:2017kvq,Belin:2018fxe,Botta-Cantcheff:2018brv,Botta-Cantcheff:2019apr,Chen:2019ror}. We will call such states $\ket{\lambda}$ and their wave-functions are given by the Euclidean path integral
\be
\braket{\varphi_0|\lambda}=\int_{\varphi(t_E=0)=\varphi_0} \mathcal{D}\varphi e^{-S_{\text{CFT}}+\int_{t_E<0} \lambda(x) O(x)} \,.
\ee
Note that the source $\lambda(x)$ should turn off sufficiently fast at $t_E=0$ such that it does not deform the theory (i.e. the Hamiltonian) but only prepares an excited state of the original theory. The operators for which we allow sources are the single-trace operators of the theory (see \cite{Haehl:2019fjz} for a discussion of their multi-trace counterpart).

\begin{figure}
\begin{center}
\includegraphics[width=0.8\textwidth]{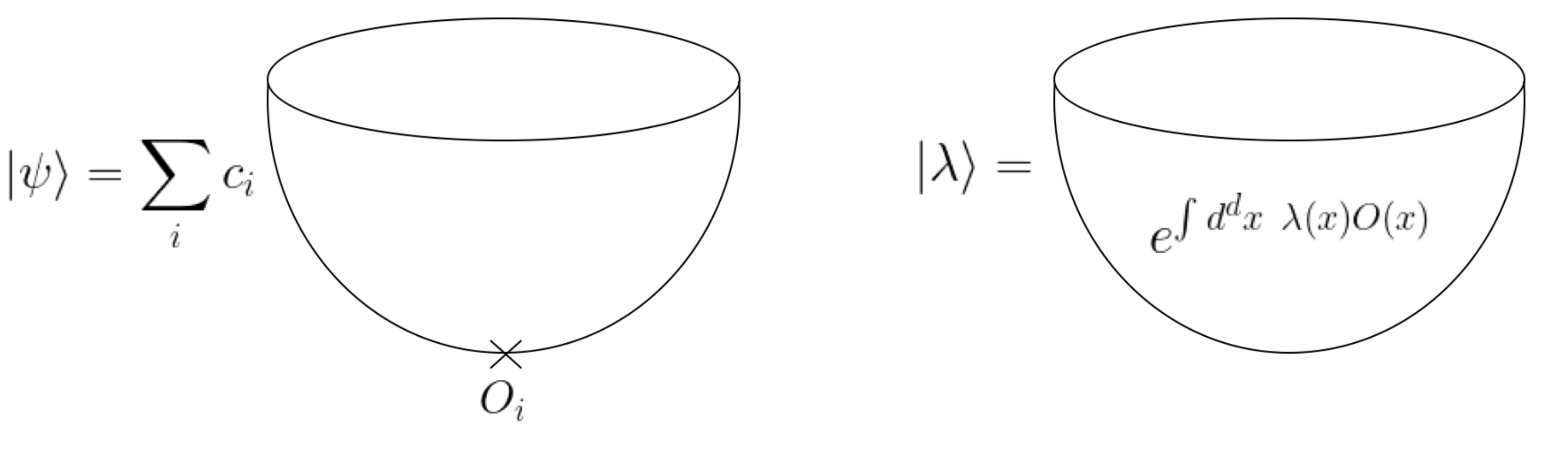}

\end{center}
\caption{The two different constructions to create excited states in a CFT. \textbf{Left}: the usual state-operator correspondence. Any state of the theory can be created by taking linear combinations of the energy eigenstates which are obtained by inserting an operator at the south pole. \textbf{Right}: a state prepared by a Euclidean path integral with a source turned $\lambda(x)$ turned on for the operator $O$. Provided such a state makes sense, it could also be written as a superposition of energy eigenstates, in a complicated way.}
 \label{euclvsstateop}
\end{figure}

We would like to emphasize the difference between the usual state-operator correspondence and these states. For the state-operator correspondence, one does not turn on a source for an operator but simply inserts it in the path integral. The operator is thus not exponentiated, which plays a crucial difference. The energy eigenstates corresponding to an insertion of a single-trace operator at the south-pole are not coherent states but rather correspond to one-particle states in the bulk perturbative quantum theory \cite{Belin:2018juv,Belin:2019mlt} and they behave very differently from coherent states. Note however that they can be obtained by the path-integral states by taking functional derivatives with respect to the sources \cite{Christodoulou:2016nej}.
We show the difference between the two types of states in Fig. \ref{euclvsstateop}. 

One can in principle try to define Euclidean path integral states in any CFT, but one needs to deal with products of operators that appear once the exponential is expanded. It remains unclear whether such states make sense for arbitrary CFTs, and for which choice of source/operator they do. As we will see, AdS/CFT hints that for holographic large $N$ CFTs the states do make sense, at least for certain class of operators.

In AdS/CFT, the standard dictionary \cite{Gubser:1998bc,Witten:1998qj} states that turning on Euclidean sources for single-trace operators corresponds to looking for a smooth solution of the gravitational equations of motion with appropriate boundary conditions for the bulk fields, as to match the sources. From the bulk point of view at the classical level, this is a Euclidean boundary value problem for an elliptic PDE (see for example \cite{Witten:2018lgb} for a discussion of this question).\footnote{Note that the fact that the PDE is elliptic does not guarantee the existence or uniqueness of a solution.}

\begin{figure}
\begin{center}
\includegraphics[width=0.5\textwidth]{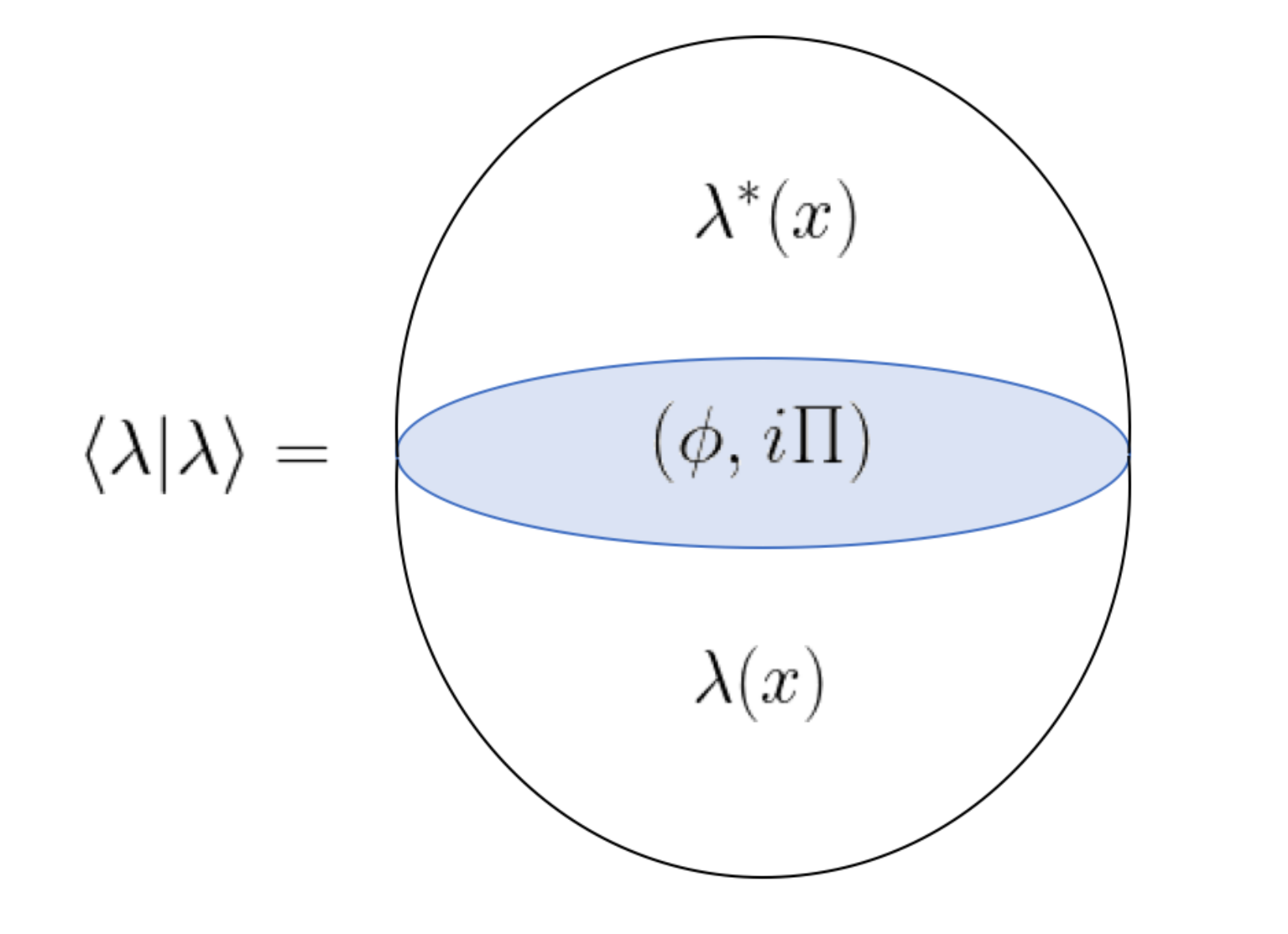}

\end{center}
\caption{The overlap $\braket{\lambda | \lambda}$, which is a Euclidean path integral in the CFT. There is a bulk geometry dual to these sources, and a time-slice where the field is purely real and the canonical momentum (or normal derivative) purely imaginary. The Lorentzian initial data is obtained by analytically continuing the geometry on this slice $(\phi,i\Pi)\to(\phi, \Pi)$.}
 \label{overlap}
\end{figure}

To read off the phase space variables $(\phi,\Pi)$ from the sources, one considers the overlap $\braket{\lambda | \lambda}$ and finds the appropriate smooth geometry. Note that it is important that we allow for complex sources, which can be seen from parameter counting: $(\phi,\Pi)$ are 2 real functions of $d$ dimensional coordinates, which maps to the real and imaginary parts of $\lambda(x)$. The dual states $\bra{\lambda}$ are obtained by inserting the conjugate sources $\lambda^*(x)$ in the northern hemisphere. The initial data is obtained by finding the $Z_2+C$ symmetric slice in the bulk, where the geometry can be analytically continued to Lorentzian signature such that the phase space variables are real. This is illustrated in Fig. \ref{overlap}. Correlation functions in Lorentzian time can then be computed using an appropriate time contour and the gluing between Euclidean and Lorentzian geometries \cite{Skenderis:2008dg}.

The mapping between boundary sources and initial data also persists at the level of the symplectic structure: the symplectic form on the classical phase space of gravitational configurations is dual to a CFT symplectic form obtained from the Fubini-Study metric pulled-back to the space of Euclidean path integral states \cite{Belin:2018fxe}. In the CFT symplectic form, VEVs and sources are canonically conjugate, as already noted in the early days of AdS/CFT \cite{deBoer:1999tgo,Papadimitriou:2004ap}. It is worthwhile to note that as for usual coherent states in quantum mechanics, these states are expected to span the Hilbert space but are over-complete. There will therefore be a non-zero (but exponentially small) overlap between distinct coherent states. This is true even for geometries that look very different, such as a thermofield double state that is dual to a black hole and two disconnected copies of AdS \cite{Jafferis:2017tiu}.

The arguments described above give strong evidence that given some sources, we can find the corresponding Lorentzian initial data and we thus have a map
\be
\ket{\lambda} \longrightarrow \ket{\phi,\Pi} \,.
\ee
The goal of this paper is to investigate whether the converse is true: given initial data, can it always be described by a state prepared by a Euclidean path integral with sources
\be
\ket{\phi,\Pi}  \xrightarrow{???}  \ket{\lambda}  \,.
\ee
This direction is perhaps even more important: We would like to understand the CFT states that describe arbitrary initial data and if for some reason this cannot be accomplished by Euclidean path integrals, it presents some challenges for AdS/CFT.

Some initial steps towards answering this question were undertaken in \cite{Marolf:2017kvq,Chen:2019ror}. The primary concern of the authors was regarding initial data that is very localized on the Cauchy slice. They found that localized initial data leads to divergent CFT sources. In this paper, we will take a slightly different approach and will argue that in fact the problem is ill-posed to start with, which will immediately raise conceptual questions that we will try to address. Finally, note that states prepared by Euclidean path integrals have proven very useful in holography, in particular in the context of holographic entanglement (see for example \cite{Faulkner:2017tkh,Balakrishnan:2020lbp}). This gives extra motivation for understanding them.

In this paper, we will consider the simplest possible scenario: a free scalar field in AdS, and we will mostly work in the limit where back-reaction is negligible. For a given mass of the scalar field, the equations of motion reduce to the two-dimensional Laplace equation, for which well-known theorems immediately imply that the initial data $\to$ source problem is ill-posed. As we will show, the initial data is not generic but rather has to be analytic.

Moreover, to obtain arbitrary analytic initial data, we will show that one must include sources in the bulk. The locations of the sources correspond to singularities in the bulk where the equation of the motions for the scalar are no longer satisfied. This is reminiscent of electrostatics where charges are required to obtain arbitrary electric field on a plane. In order to tell if a given initial data corresponds to a smooth Euclidean section or rather leads to singularities, we formulate a criterion that distinguishes between the two scenarios. This criterion involves an integral of the data on the Cauchy slice and is therefore not a local condition. For meromorphic initial data, it can be evaluated by a contour integral in the complex plane and depends only on the residues of the initial data function at its poles.
Finally, we briefly speculate on the physical interpretation of these singularities both in the bulk and in the CFT. As we go along, we will illustrate our results with a series of concrete examples. 

The paper is organized as follows: in section \ref{posedness}, we present our simple model and recall theorems for the Laplace equation that imply that the initial data $\to$ source problem is ill-posed. In section \ref{examples}, we give a series of example including bulk solutions with or without sources, and briefly comment on backreaction. In section \ref{constraining} we give an integral equation that initial data must satisfy in order for the bulk to be regular. In section \ref{discussion}, we discuss the implications of our results for the dictionary of AdS/CFT and comment on possible connections between the singularities needed in the bulk and UV objects such as D-branes.

\section{Initial data to Euclidean sources: ill-posedness \label{posedness}}
To demonstrate our results in a simple and concrete setting we consider Einstein gravity in the bulk, minimally coupled to a scalar field,
\be
S =  \int d^{d+1}x \sqrt{|g|}\left[\frac{1}{16\pi G_N}\left(R-2\Lambda\right) - \frac{1}{2}(\partial\phi)^2-\frac{1}{2}m^2 \phi^2\right],\label{action}
\ee
with cosmological constant $\Lambda = -\frac{d(d-1)}{2L^2}$ and scalar mass $m^2= \Delta(\Delta -d)/L^2$. 
In the probe limit where the backreaction of $\phi$ is neglected, the scalar equation of motion becomes the Laplace equation on half of $\mathbb{R}^{d+1}$ for certain choices of $\Delta$. This will enable us to import known results for the well-posedness of the Cauchy problem for the Laplace equation, where we focus on the existence of solutions.

To see this explicitly, consider the Poincar\'e metric together with the following field redefinition for $\phi$,
\be
ds^2 = \frac{L^2}{z^2}\left(dz^2 + d\vec{x}^2\right)\qquad \phi(z,\vec{x}) = z^{\frac{d-1}{2}} f(z,\vec{x})
\ee
then provided $\Delta = (d \pm 1)/2$,  $f(z,\vec{x})$ obeys the following equation of motion in the probe limit,
\be
(\partial_z^2 + \delta^{ij}\partial_i \partial_j)f(z,\vec{x}) = 0. \label{laplace}
\ee
In other words, $f$ obeys the Laplace equation on half of $\mathbb{R}^{d+1}$, $z>0$, where $z=0$ corresponds to the boundary of AdS$_{d+1}$. If we take $\Delta = (d + 1)/2$ then we see that the CFT source function is $\lambda(x) = f_{z= 0}$ and the remaining data $\partial_z f_{z = 0}$ determines the operator VEV $\left<O\right>$ after performing holographic renormalisation \cite{Bianchi:2001kw}. For example, in the absence of sources $\lambda = 0$ we have $\left<O\right> = -\partial_z f_{z = 0}$ \cite{deHaro:2000vlm}. For $\Delta = (d - 1)/2$ the source and VEV identifications are exchanged.

To specify a Cauchy problem we single out one of the boundary coordinates, $\tau \equiv x^1$, then the associated initial data is $u = f(\tau = 0)$ and its normal derivative, $v = f_{\tau} (\tau = 0)$. To show that this problem is ill-posed we need only consider the following theorem.

\begin{figure}[h!]
\begin{center}
\includegraphics[width=0.65\columnwidth]{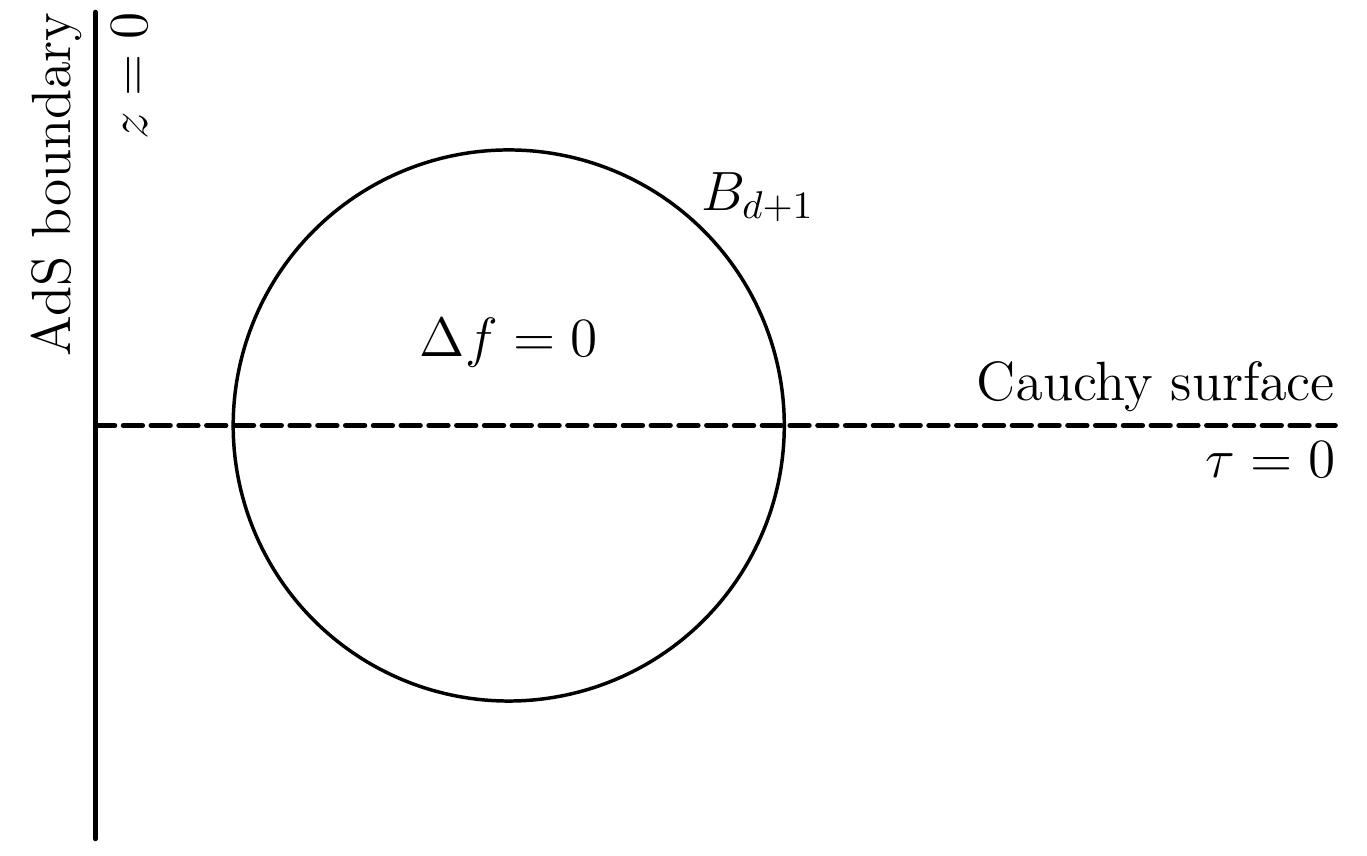}
\caption{Any region of the initial data surface is contained within some ball $B_{d+1}$. Following Theorem 1, the solution must be analytic inside this ball, which in particular includes the initial data.}
\label{ball}
\end{center}
\end{figure}

\begin{theorem}[analyticity]
Consider an open region $\Omega$. Any $f\in C^2(\Omega)$ solving $\Delta f = 0$ in $\Omega$ is analytic in $\Omega$.
\end{theorem}
The proof can be found in textbooks on PDEs, see for example section 2.2 theorem 10 (analyticity) in \cite{evans2010partial}.\footnote{By analytic, we mean \textit{real analytic} which implies a function admits a Taylor expansion around any point of the open region $\Omega$ and has a finite radius of convergence.} To apply it to our Cauchy problem, consider a ball that includes a portion of the $\tau = 0$ surface, e.g. as in Figure \ref{ball}.
%%%%
%%%%
Then by the above theorem, $f$ must be analytic everywhere in the ball, including a portion of the data on the $\tau = 0$ surface. Thus generic initial data in $C^2(\Sigma)$ fails to solve the Laplace equation, and so the problem is ill-posed. Instead one must restrict to analytic solutions in the bulk with analytic initial data. Note however that this does not imply analytic $\lambda(\vec{x})$ since it does not belong to any open region where a solution is required.\footnote{Note that if $\tau=0$ is treated as a boundary (which is not the case here), the problem remains ill-posed. There are strong constraints placed on the compatibility of the initial data $f,f_{\tau}$ such that there is a solution in $\tau>0$ in the neighbourhood of $\tau=0$ \cite{hadamard1902}.} This latter point is sharply illustrated by the Janus solutions \cite{Bak:2003jk} where $\lambda$ is non-analytic at the defect, but the extension into the bulk is regular.

Given analytic initial data $f,f_{\tau}$ it follows from the Cauchy-Kovalevskaya theorem  that a unique analytic solution can be constructed in the neighbourhood of any point on the $\tau = 0, z>0$ surface. However, this theorem does not guarantee that there is a solution for all $\tau$ and $z>0$; only up to some radius of convergence. Physically we will see in our examples (section \ref{examples}) that this breakdown occurs at singularities which indicate locations in the bulk where delta-function sources have been turned on. Nevertheless this is not an obstruction  to extracting the corresponding $\lambda$, since unique analytic solutions can be obtained on $\mathbb{R}^{d+1}$ minus these singular points.

Finally it is worth emphasising that the problem is ill-posed also because solutions are not continuously dependent on initial data. This is exemplified by Hadamard's example \cite{Hadamard:231266}; the initial data $f  = \sin{(k z)}/k,\; f_\tau = 0$, corresponds to the unique analytic solution
\be
f = \frac{1}{k} \sin{kz} \cosh{k \tau}.
\ee
This illustrates that high-$k$ modes in the initial data lead to terms which diverge rapidly in Euclidean time. In particular, as $k\to\infty$ the initial data becomes arbitrarily close to that of the trivial solution $f=0$, however, at any finite $\tau\neq0$ the solution deviates from the trivial solution as $\sim e^{k\tau}/k$.

\section{Analytic initial data: examples\label{examples}}
In the preceding section we established that the Cauchy problem for $C^2$ initial data is ill-posed, finding that every solution must be analytic in the bulk with analytic initial data. In the following section \ref{eg:trans}  we consider examples of such analytic solutions, and show that they contain singularities in the bulk. This shows that the class of initial data prepared in the usual way by a boundary value problem subject to bulk regularity only realises a subset of all possible analytic initial data. A natural consequence of this is the existence of singular solutions with $\lambda = 0$ and $\left<O\right>\neq 0$, which we term pure VEV solutions constructed in section \ref{purevev} with fully backreacted variants constructed in section \ref{backreaction}.

\subsection{A bulk singularity \label{eg:trans}}
We first look at solutions possessing translational invariance in $d-1$ directions of AdS$_{d+1}$. Such solutions are governed by the two dimensional Laplace equation. We further restrict to time-symmetric solutions for simplicity, so that $f_\tau(z,\tau = 0) = 0$. Given the remaining initial data $f(z,\tau = 0) = u(z)$ the unique analytic solution is obtained,
\be \label{generalsolution}
f(z,\tau) = \frac{u(z + i \tau) + u(z-i \tau)}{2}.
\ee
For example the initial data,
\be 
u(z) = \frac{z}{\tau_0^2+(z-z_0)^2},\label{eginitial}
\ee
corresponds to the solution
\be
f(z,\tau) = \frac{z^3  - 2 z^2 z_0 + z(\tau^2 + z_0^2 + \tau_0^2) - 2 z_0 \tau^2}{(\tau^2+(z-z_0)^2)^2  +  2(z-z_0-\tau)(z-z_0+\tau)\tau_0^2 + \tau_0^4}.\label{egsol}
\ee
This has singularities at $\tau = \pm \tau_0, z = z_0$ and one can verify that the Laplace equation is not solved at these locations, instead this solves the Poisson equation with delta-function sources on right hand side of \eqref{laplace}, here arranged into dipoles. Introducing $\vec{y} = (z,\tau)$ this is,
\be
-\Delta_y f = \lim_{\mu \to 0} \frac{q}{2\mu} \left(\delta^{(2)}(\vec{y} - \vec{y}_0 - \mu \hat{n})-\delta^{(2)}(\vec{y} - \vec{y}_0 + \mu \hat{n})\right) + (\tau \to -\tau),
\ee
with dipole angle $\hat{n} = \frac{1}{\sqrt{z_0^2 + \tau_0^2}}(-\tau_0,z_0)$ and charge $q = -\pi \frac{\sqrt{z_0^2 + \tau_0^2}}{\tau_0}$. We have thus uncovered non-perturbative objects in the theory realised as sources on the right hand side of our bulk equations. This is akin to the role played by the electron in solutions to electrostatics, or D-branes in the solutions to $10$-dimensional supergravity. 

Generalising these solutions, it is clear that by placing an arbitrary source distribution $\rho(z,\vec{x})$ on the right hand side of \eqref{laplace}, 
\be
-(\partial_z^2 + \delta^{ij}\partial_i \partial_j)f(z,\vec{x}) = \rho(z,\vec{x}), \label{poisson}
\ee
convolving $\rho(z,\vec{x})$ with the fundamental solution to the Laplace equation provides the unique solutions for an infinite class of analytic initial data provided there are no sources placed on the initial data surface, $\rho(\tau = 0) = 0$.\footnote{The cohomogeneity-2 examples above correspond to smearing the fundamental solution over $d-1$ planes.} 

\subsection{Pure VEV solutions}\label{purevev}
Continuing the example of \eqref{egsol} we can read off the source at the boundary,
\be
\lambda = -\frac{ 2 z_0 \tau^2}{z_0^4 + 2 z_0^2(\tau^2 + \tau_0^2) + (\tau^2-\tau_0^2)^2},
\ee
and we can construct a second solution with this same boundary source, but requiring regularity in the interior $z>0$. This is a well-posed boundary value problem. Once obtained, we can subtract it from \eqref{egsol} to obtain a solution with the same singularity structure for $z>0$ but with $\lambda = 0$. We shall refer to this as a pure VEV solution. Denoting the solution in \eqref{egsol} as $f_{\eqref{egsol}}$, in this case it is explicitly,\footnote{This simple form of the solution arises because $f_{\eqref{egsol}}$ just happens to have no singularities for $z<0$. More generally it will not take this form.}
\be
f(z,\tau) = f_{\eqref{egsol}}(z,\tau)- f_{\eqref{egsol}}(-z,\tau).
\ee
Note that this differs from \eqref{egsol} by the addition of singularities for $z<0$. We can view the original example \eqref{egsol} as a solution in this singular sector, further deformed by a choice of $\lambda$.

 %%%%
\begin{figure}[h!]
\begin{center}
\includegraphics[width=0.65\columnwidth]{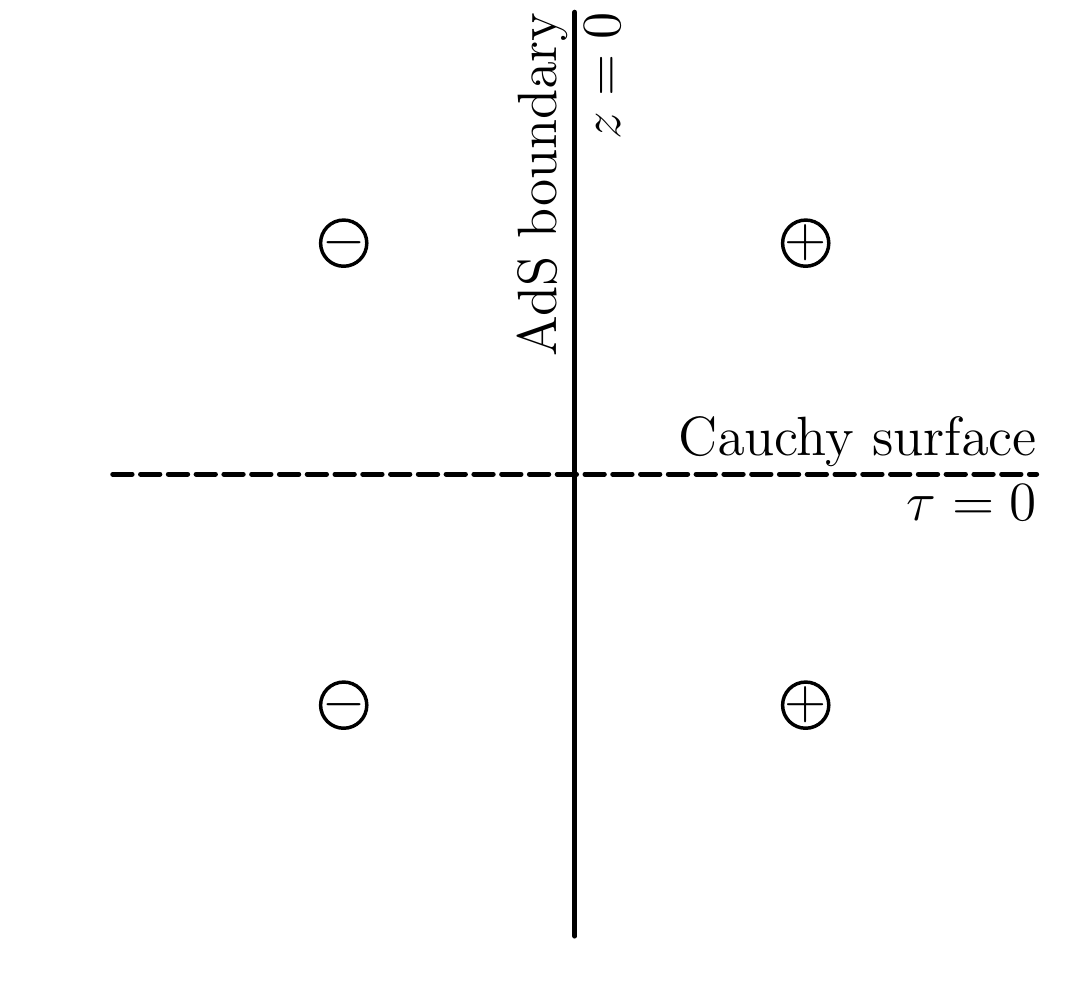}
\caption{An example construction of a pure VEV solution considered as an electrostatics problem, by the insertion of elementary point charges in the bulk $z>0$ and their images in $z<0$. The resulting scalar field profile is given in \eqref{points}. \label{planarvac}}
\label{capcoords}
\end{center}
\end{figure}
%%%%
More generally, we can arrange for solutions $\lambda(x) = 0$ as an electrostatics problem using the method of images. For example, placing single charges at $(z_0,\tau_0)$, $(-z_0,\tau_0)$, $(-z_0,-\tau_0)$, $(z_0,-\tau_0)$ with charges $+1,-1,-1,+1$ respectively, as illustrated in figure \ref{planarvac}, is the analytic solution developed from the time symmetric initial data, 
\be
f(z, \tau = 0) = \frac{1}{2\pi} \log\frac{(z+z_0)^2 + \tau_0}{(z-z_0)^2 + \tau_0^2}, \label{points}
\ee
but differs from the vacuum solution by non-zero VEVs at zero source,
\bea
\lambda  &=& 0,\\
\left<O\right>&=& -\frac{z_0}{\pi} \left(\frac{1}{z_0^2 + (\tau-\tau_0)^2}+\frac{1}{z_0^2 + (\tau+\tau_0)^2}\right).
\eea

\subsection{Incorporating backreaction \label{backreaction}}
We may construct fully backreacted examples of these new sourceless solutions by utilising the perturbative solutions as a seed, and solving the full Einstein equations subject to the condition that $\lambda = 0$. This is most straightforward when we construct spherically symmetric solutions centred on the sources. We adopt the following ansatz,
\be
ds^2 = L^2\left(\frac{dr^2}{f(r)} + r^2 d\Omega_{d}^2\right), \quad \phi = \phi(r), \label{ballansatz}
\ee
where $\phi = 0, f = 1 + r^2$ corresponds to global AdS$_{d+1}$. One can analytically solve for the metric function $f(r)$ in terms of $\phi(r)$ and its derivatives, then $\phi(r)$ obeys a second order ODE.

Based on the previous examples, we expect singularities consistent with point-sources. In particular in the bulk we  expect divergences in the probe limit that behave as
\be
\phi \propto 
\begin{cases}
\frac{1}{r^{d-1}} & d > 1\\
\log{r} & d=1
\end{cases}\label{sing}
\ee
near the singularity placed at $r=0$.
Indeed, this is the case for the ansatz \eqref{ballansatz}, where in the probe limit we have,
\be
\phi(r) = c\, r^{-\Delta}\; {}_2 F_1\left(\frac{\Delta-d+1}{2},\frac{\Delta}{2},\frac{2\Delta - d +2}{2},-\frac{1}{r^2}\right),
\ee
which diverges as \eqref{sing} near $r=0$. However, such behaviour is altered by backreaction, which we now discuss for $d>1$. Near $r=0$ the power-law divergence becomes logarithmic. This can be seen by constructing a backreacted solution as a series around $r=0$,
\be
\phi(r) =
\pm\sqrt{\frac{d(d-1)}{8\pi G_N}}\left(\log{r} + \gamma_1  - \gamma_2 r^{2(d-1)} + \ldots\right),\label{phiIR}
\ee
where $\gamma_1,\gamma_2$ correspond to undetermined coefficients in the $r=0$ expansion. The behaviour \eqref{phiIR} leads to power-law divergences in the metric and in the Ricci scalar, 
\be
f(r) = \frac{1}{\gamma_2}\frac{1}{4(d-1)}\frac{1}{r^{2(d-1)}} + \ldots, \qquad R = \frac{1}{\gamma_2}\frac{d}{4 L^2}\frac{1}{r^{2d}} + \ldots.
\ee
In general the full solution can be constructed by integrating from $r=0$ and shooting for $\lambda = 0$ at the boundary by adjusting the two pieces of data $\gamma_1,\gamma_2$ accordingly. This leads to a one parameter family satisfying $\lambda = 0$, where the remaining parameter determines the strength of the singularity. Initial data can subsequently be extracted from any choice of Cauchy surface that avoids $r=0$. Such a choice would not lead to real Lorentzian initial data, but it would be straightforward to generalize the solution (for example by having 2 singularities, one above and one below $\tau=0$) such that the data is real.

In the case $\Delta = d$ the solution can be written in closed form,
\be
f = 1 + r^2 + \frac{8 \pi G_N \alpha^2}{d(d-1)}\frac{1}{r^{2(d-1)}},\qquad \phi' = \frac{\alpha}{r^d\sqrt{f}}.
\ee
From this solution it is clear how the change from $r^{1-d}$ behaviour in the probe limit to $\log{r}$ behaviour with backreaction comes about from resumming the amplitude perturbative expansion (i.e. small $\alpha$). 

Note that while this is a solution to our bottom-up bulk model \eqref{action}, it closely resembles D-instanton solutions constructed in 10D supergravity \cite{Gibbons:1995vg, Bergshoeff:2005zf} in both structure and the quantitative nature of the logarithmic and power-law divergences. In particular the super-extremal instantons discussed in \cite{Bergshoeff:2005zf} have a bulk metric of the form \eqref{ballansatz} in $d=4$ with
\be
f = 1+r^2 + \frac{q^2}{L^6r^6}.
\ee

In the example above we chose to place the singularity at the origin of coordinates to make rotational symmetry manifest. However due to the maximal symmetry of hyperbolic space there are no physically privileged points. Naively these $d+1$ free parameters labelling the position in the bulk correspond to the $d$ positional and a single size collective coordinates of $SU(N)$ gauge theory instantons. We will discuss this further in section \ref{discussion}.

\section{Constraining initial data}\label{constraining}
We have established that analytic initial data can lead to singularities in the bulk. A natural question arises: what condition must be placed on the initial data so that there are no singularities? We know that such choices exist, since one can construct a solution as a boundary value problem specifying $\lambda(\tau)$ and imposing regularity, from which one can read off the initial data.  This is the choice that a CFT naturally makes; given a source $\lambda$ the dynamics of the CFT determine the VEVs, which maps into a particular set of initial data. In this section we obtain an integral equation that the initial data must obey so that such dynamical constraints are met.

\begin{figure}
\begin{center}
\includegraphics[width=0.7\textwidth]{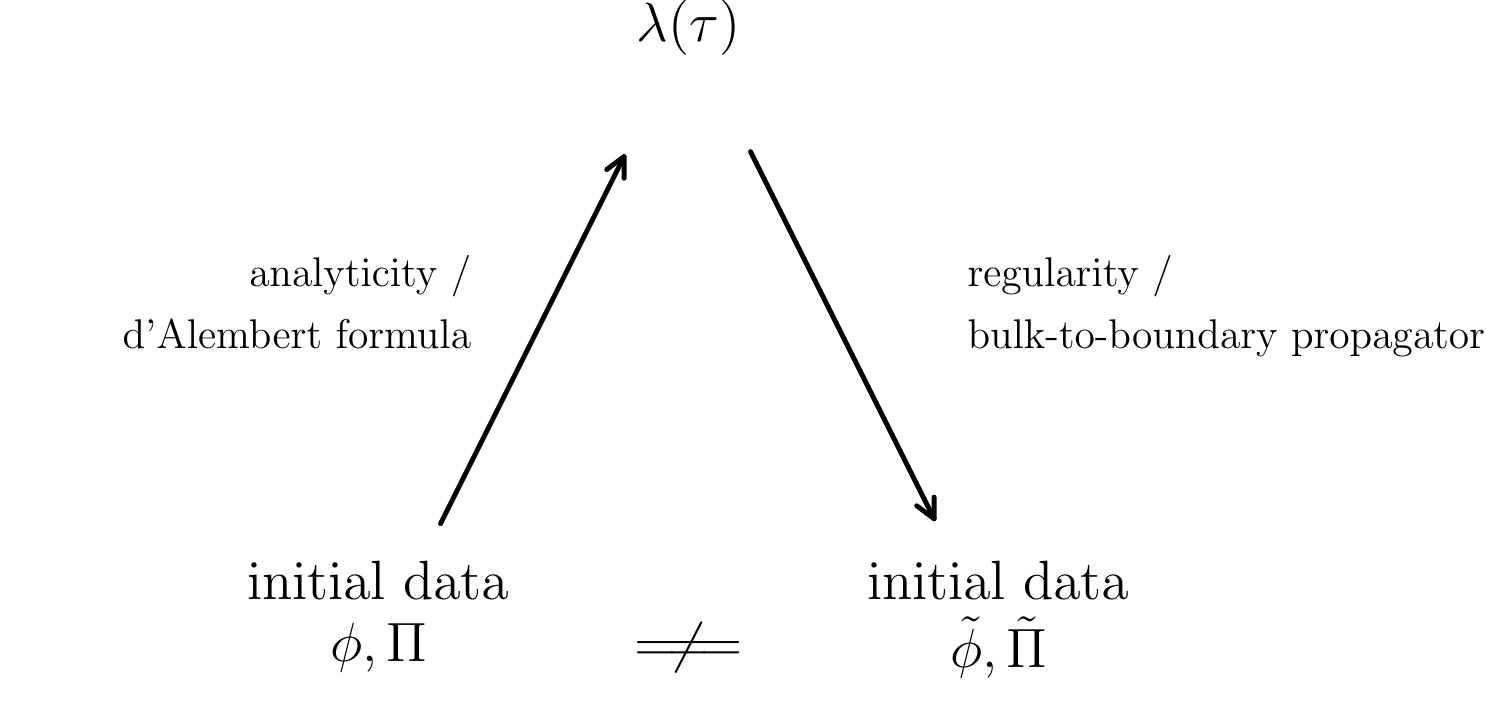}
\end{center}
\caption{An \emph{aide-visuelle} to the discussion of maps between sources and analytic initial data. \textbf{Left arrow}: to obtain a solution from initial data one can use the appropriately analytically continued d'Alembert formula, and read off $\lambda, \left<O\right>$. This procedure generically results in bulk singularities. \textbf{Right arrow}: Alternatively, one can start with $\lambda$ and solve given bulk regularity, and read off initial data and $\left<O\right>$. Generically this results in different analytic initial data and $\left<O\right>$ for the same $\lambda$. In a probe limit, subtracting $\phi,\Pi$ from $\tilde{\phi},\tilde{\Pi}$ gives what we have termed a pure VEV solution, namely a solution with $\lambda = 0$ and $\left<O\right> \neq 0$ which generically contain singularities in the bulk (note such solutions also exist including backreaction, see section \ref{backreaction}).}
\label{noncom}
\end{figure}

The initial data corresponding to a regular solution $\tilde{u}(z)$ can be constructed using the bulk-to-boundary propagator on a mode-by-mode basis,
\be \label{inverse}
\tilde{u}(z) =  2\int_0^\infty d\omega \hat{\lambda}(\omega) e^{-\omega z}\quad \text{where} \quad    \hat{\lambda}(\omega) = \frac{1}{2\pi} \int_{-\infty}^{\infty}  d\tau \lambda(\tau) e^{i \omega \tau},
\ee
where we restrict to time-reversal invariant solutions for simplicity. On the other hand, given an arbitrary choice of time-symmetric initial data $u(z)$ we can read off $\lambda$ using the analytically continued d'Alembert formula \eqref{generalsolution}. These distinct maps are summarised in figure \ref{noncom}.

Thus the condition of a regular bulk translates to a simple condition on initial data $u(z)$ by first applying the inverse analytic map, then the forward regular-propagator map, and demanding that the resulting $\tilde{u} = u$. In other words, a necessary and sufficient condition for a regular solution is $I(z)=u(z)$ where
\be
I(z) \equiv \frac{1}{\pi} \int_0^\infty d\omega \int_{-\infty}^{\infty} d\tau \frac{u( i \tau) + u(-i \tau)}{2} e^{-\omega z+i \omega \tau}.
\ee
The $\omega$-integral can be performed for all $\text{Re}(z)>0$,
\be
I(z) = \frac{1}{\pi}\int_{-\infty}^{\infty} d\tau \frac{1}{z-i \tau}\frac{u( i \tau) + u(-i \tau)}{2}. \label{integral}
\ee
To evaluate this integral we assume that the initial data function $u(i \tau)$ obeys appropriate boundedness conditions so the integral along the real $\tau$ axis can be extended to a contour in the complex $\tau$ plane that closes either the LHP or UHP at our discretion. Choosing the contour in the UHP means that the pole at $z = i \tau$ does not contribute. 

Some examples where we can draw conclusions from \eqref{integral} are as follows. If $u(z)$ is an odd function, then $I=0$ and hence the solution is singular. For meromorphic $u(z)$ the condition (28) becomes a detailed condition on the residues, for example, consider the initial data \eqref{eginitial}, $u(z) = z/(\tau_0^2+(z-z_0)^2)$ for $z_0>0$ (which we know to be singular from section \ref{eg:trans}), as well as the initial data
\be
u(z) = -\frac{z}{\tau_0^2+(z+z_0)^2}\label{egreg}
\ee
 (which corresponds to a regular solution). Both have the same source function $\lambda(\tau)$ and also the same $I(z)$, \eqref{integral}. Closing \eqref{integral} in the UHP gives residues coming from poles at $\tau = \pm \tau_0 + i z_0$, resulting in
\be
I(z) = -\frac{z}{\tau_0^2+(z+z_0)^2},
\ee
which is equal to the regular case initial data, \eqref{egreg}. Thus the former data \eqref{eginitial} does not satisfy the condition but the latter data \eqref{egreg} does.

To conclude, we have seen in this section that for generic initial data, the prescription \rref{inverse} is not the inverse of the d'Alembert map. We would now like to comment on a different approach taken in \cite{Marolf:2017kvq} where the boundary sources were expressed in terms of the initial data by formally inverting a relation constructed using bulk-to-boundary propagators in momentum space (equation (37) of \cite{Marolf:2017kvq}). The authors left it as an open question as to whether the inverse map proposed is well-defined.  We have not been able to directly evaluate the inverse Laplace transform, but the best case scenario is that it reproduces the d'Alembert map. In any case, we have seen that the regular bulk-to-boundary propagator cannot be the inverse of their (37).

\section{Discussion \label{discussion}}

In this paper, we have discussed the relation between CFT states prepared by a Euclidean path integral and coherent states of the dual bulk theory which are parametrized by a choice of initial data. We considered a simple model consisting of Einstein gravity with a minimally coupled scalar in AdS. While a class of initial data can be obtained by specifying the sources in the Euclidean section and solving the bulk equations of motion demanding regularity, we have argued that the converse is not true: specifying the initial data and trying to read off the sources is not a well-posed problem. Instead, one must restrict to analytic solutions up to bulk singularities. As a corollary, we have shown that initial data with a non-singular bulk Euclidean section are measure-zero in the set of possible initial data. In our simple model, we have shown that arbitrary analytic initial data can only be obtained by including bulk singularities, which can be interpreted as delta-function sources. We now discuss some open questions.

\subsection{Bulk sources and non-perturbative objects}

We have seen in our simple scalar model that bulk singularities must be included in order to obtain arbitrary initial data. In this section, we discuss possible interpretations for these singularities. 
As discussed in section \ref{backreaction}, similar backreacted singular solutions have appeared before in the literature as D-instantons (see for example \cite{Bergshoeff:2005zf}). It would therefore be tempting to interpret the presence of the bulk sources in this same way. 
 However, this presents some complications: first, since the CFT is usually a gauge-theory, an instanton describes a transition amplitude between two states of different winding number \cite{Callan:1976je, PhysRevLett.37.172,  PhysRevD.17.2717}. This would suggest that not all classical configurations correspond to overlaps between a state and itself, but one may need to consider transition amplitudes. This could in principle mean that not all initial data corresponds to a state \`a la \rref{energybasis}. Second, one typically sums over all instantons rather than picking a particular instanton sector. The initial data would then correspond to exponentially suppressed corrections to the wave-function, rather than a leading contribution. It would be interesting to understand how to project to a particular instanton sector. We hope to return to these questions in the future.

In general, the bulk sources may not need be D-instantons, but could correspond more generally to other types of D-branes. In a top-down construction, the dimensionality of the D-brane may depend on the internal manifold at hand, which we have not discussed here. It would be interesting to probe this issue further in a model such as $\mathcal{N}=4$ SYM. Note however that the phase-space interpretation of the internal manifold is currently not understood for the path-integral states considered here, and this would need to be worked out first (see \cite{Skenderis:2006uy} for steps in this direction). One may also hope that the singularities are resolved in the full string theory, for example by turning on stringy modes. This could imply that the VEVs for the stringy operators would be non-zero, and likewise that they would have a non-zero profile at the initial data slice. Such considerations may also be probed using a top-down model.

Taking a step back, one may wonder why non-perturbative objects are needed in the Euclidean section when the corresponding Lorentzian section appears to remain in the low-energy EFT. While we do not have a definitive answer to this question, we offer some speculative comments. We may draw an analogy with a simple problem of undergraduate physics: electrostatics. In electrostatics, if we wish to solve for the most arbitrary electric field on a plane, we will quickly discover that we need to allow for the presence of electrons. In our setup, we have half of the Euclidean plane, and the objects in $z<0$ can be replaced by a boundary condition at $z=0$; this is the role played by the AdS boundary. However electrons are still required in $z>0$ in order to obtain an arbitrary electric field. The parallel with gravity seems straightforward, by looking for arbitrary initial data we require the existence of D-branes.

Finally it is interesting to contrast these singularities with instabilities that appear in the model \eqref{action} in Lorentzian evolution \cite{Bizon:2011gg}. This instability requires interactions, while the Euclidean singularities are already seen at the probe level.

\subsection{Approximating initial data by truncating in momentum space}

Given the presence of bulk singularities, one may be concerned with UV behaviour. A natural suggestion is to construct an approximated set of initial data by tampering with short wavelengths such that a regular bulk configuration is obtained. A simple example of this would be to employ a basis of modes, such as dilatation eigenfunctions in global AdS, and keep only finitely many of them. Such a procedure would manifestly yield a regular bulk configuration (which is also analytic). Some arguments for this perspective is that we should not take modes of Planckian frequencies too seriously in the first place, so it is natural to truncate high frequency modes. Moreover, a mode truncation enables a direct interpretation of the state following equation \rref{energybasis}. However, as may be anticipated given the ill-posedness of the problem (and in particular Hadamard's example illustrating the sensitivity of solutions to high-$k$ data discussed in section \ref{posedness}), this perspective is not particularly useful, and a more natural construction results from allowing the existence of bulk singularities.

We will give three main arguments for this perspective. First, consider again the initial data \rref{eginitial} that leads to a singular bulk and decompose it into Fourier modes in $z$, with a maximal wavevector $k_{\max}$, after which the modes are discarded. This initial data can be constructed by convolving $u(z)$ with $\text{sinc}(k_{\max} z)$. One can then obtain the boundary sources for this data using \rref{generalsolution}. For $\tau_0 = z_0 = 1$ this is given by,
\be
\lambda(\tau) = -\frac{\tau^2}{4+\tau^4} + e^{k_{\max}(\tau-1)} \frac{\tau \cos{k_{\max}} - (\tau-2)\sin{k_{\max}}}{2(2-2\tau+\tau^2)}+(\tau \to -\tau).\label{bandwidthsource}
\ee
In the limit where the cutoff $k_{\max}$ is removed, $k_{\max}\to\infty$, the second term in \eqref{bandwidthsource} vanishes provided $\tau < 1$ and diverges otherwise. This can be easily seen from the exponential in the second term of \eqref{bandwidthsource}. Thus \eqref{bandwidthsource} only provides an approximation valid in the strip $|\tau| < 1$.\footnote{Recall that the exact solution has singularities in the bulk at $|\tau| = z = 1$, which is why $|\tau| = 1$ is singled out here.} In this region the source function of the exact solution is recovered in the limit. In this sense, the cutoff solution only approximates the exact solution near the initial data surface, in particular not at arbitrary points of the boundary where we want to read off the sources. 

Secondly, notice that the bandwidth-limited solutions have sources \eqref{bandwidthsource} which diverge exponentially as $\tau\to \pm \infty$, with a rate $k_{\max}$. 
Similarly, a decomposition into finitely many dilatation eigenfunctions of global AdS results in a solution that diverges exponentially in Euclidean time, with the fastest growth rate set by the highest mode kept. The global case reveals a vanishing source function except for a set of delta functions and derivatives thereof acting on the north or south pole. In either case, it is not clear why a finite sum of such divergent terms is a useful description, particularly since they depend strongly on the arbitrary cutoff chosen. 

Finally, any approximation that produces vastly different sources would destroy the nice duality between the bulk symplectic structure on the classical phase space and the boundary symplectic form \cite{Belin:2018fxe}. 
While it is not completely clear to us how to include bulk singularities in the duality between symplectic forms, there is at least some hope that it can be done, and we could still discuss nearby solutions that have the same number of bulk singularities. This will most likely require considering super-selection sectors for different number of singularities. On the other hand, any tampering with the large frequencies would drastically affect the sources and destroy the associated phase spaces: nearby solution in terms of initial data may have large deviations in their corresponding sources. The duality between the bulk and CFT symplectic forms and associated phase-spaces is a useful organization to understand the structure of semi-classical states, which gives extra motivation for considering bulk singularities rather than approximate solutions.

\section*{Acknowledgements}
It is a pleasure to thank N. Bobev, R. Emparan, N. Engelhardt, C. Gundlach, G. Horowitz, D. Marolf, H. Maxfield, C. Rabideau, G. Sarosi, K. Skenderis, J. Sonner, M. van Raamsdonk, S. Vandoren and T. Wiseman for helpful discussions. Additionally B.W. would like to extend a special thanks to the Trauma and Orthopaedics Unit at University Hospital Southampton, as well as the Odstock Ward in Salisbury District Hospital for their hospitality.
The work of A.B. is supported in part by the NWO VENI grant 680-47-464 / 4114.
The work of B.W. is supported by a Royal Society University Research Fellowship.

\bibliographystyle{ytphys}
\bibliography{ref}

\end{document}